\documentclass[12pt]{article}
\usepackage{epsfig}
\usepackage{psfrag}
\usepackage{latexsym}
\usepackage{amssymb}
\usepackage{amsmath}
\usepackage{mathrsfs} 
\def\be{\begin{equation}}
\def\ee{\end{equation}}
\def\bc{\begin{center}}
\def\ec{\end{center}}
\def\bea{\begin{eqnarray}}
\def\eea{\end{eqnarray}}

\newcommand{\ba}{\begin{array}{c}}
\newcommand{\bad}{\begin{array}{ccc}}
\newcommand{\ea}{\end{array}}

\newcommand{\tetaot}{\mbox{$\theta_{13}$}}
\newcommand{\tetatt}{\mbox{$\theta_{23}$}}

\newcommand{\delot}{\mbox{$\Delta_{atm}$}}

\def\nn{\nonumber}

\begin{document}

\begin{titlepage}
\vspace*{-1cm}
\hfill{RM3-TH/12-2}
\vskip 2.5cm
\begin{center}
{\Large\bf Revisiting the T2K data using different models for the neutrino-nucleus cross sections}
\end{center}
\vskip 0.2  cm
\vskip 0.5  cm
\begin{center}
{\large D. Meloni}~\footnote{e-mail address: meloni@fis.uniroma3.it} \\
\vskip .1cm
Dipartimento di Fisica ``E.~Amaldi'', Universit\`a degli Studi Roma Tre, Via della Vasca Navale 84, 00146 Roma, Italy
\\
\vskip .2cm
{\large M. Martini}~\footnote{e-mail address: mmartini@ulb.ac.be}
\\
\vskip .1cm
Institut d'Astronomie et d'Astrophysique, CP-226, Universit\'e Libre de Bruxelles, 1050 Brussels, Belgium
\end{center}
\vskip 0.7cm
\begin{abstract}
\noindent
We present a three-flavour fit to the recent $\nu_\mu \to \nu_e$ and $\nu_\mu \to \nu_\mu$ T2K oscillation data with different models for the neutrino-nucleus cross section.
We show that, even for a limited statistics, the allowed regions and best fit points in the $(\theta_{13},\delta_{CP})$ and $(\theta_{23},\Delta m^2_{atm})$ planes
are affected if, instead of using the Fermi Gas model to describe the quasielastic cross section,
we employ a model including the multinucleon emission channel.
\end{abstract}
\end{titlepage}
\setcounter{footnote}{0}
\vskip2truecm

\noindent
\section{Introduction}
\label{intro}
Recently the T2K collaboration has released data in both $\nu_\mu\to\nu_e$ appearance \cite{Abe:2011sj}  and $\nu_\mu\to\nu_\mu$ disappearance \cite{giganti} modes; in the first case, 
six events passed all the selection criteria, implying (under the assumption of a normal ordering of the neutrino mass eigenstates):
\bea
\sin^2(2\theta_{13})_{T2K}=0.11 \,,
\eea  
with the CP phase $\delta_{CP}$ undetermined.
In the disappearance channel, the  31 events collected by T2K are fitted with:
\bea
(\sin^2 2\theta_{23})_{T2K} =0.98  \qquad    |\Delta m^2_{atm}|_{T2K} = 2.65 \cdot 10^{-3} \,\text{eV}^2\nn\,.
\eea
The aim of this work is to reanalyse the T2K data to assess the impact of different 
models for the  $\nu$-nucleus cross sections on the determination of oscillation parameters. 
This work can be considered as a generalization of Ref. \cite{FernandezMartinez:2010dm}, 
where the impact of different modelizations of quasielatic cross sections in the low-gamma beta-beam regime was analyzed.
In the present case we consider two different models involving not only quasielastic but also pion production and inclusive cross sections. 
On one hand, we choose a model as similar as possible to the one used by the T2K collaboration.
They  simulate the neutrino-nucleus interaction using the NEUT Monte Carlo Generator \cite{Hayato:2002sd}. 
Even if we do not know the details of the last tunings performed by the collaboration to take into account for the recent measurements of K2K \cite{Gran:2006jn,:2008eaa}, 
MiniBooNE \cite{AguilarArevalo:2009eb,AguilarArevalo:2010zc} and SciBooNE \cite{Kurimoto:2009wq,Nakajima:2010fp}, we treat the several exclusive channels 
using the same models implemented in NEUT. 
As a consequence, we consider the Fermi Gas \cite{Smith:1972xh} for the quasielastic channel and the Rein and Sehgal model \cite{Rein:1980wg} for pion production. 
The second model considered in our analysis is the one of Martini, Ericson, Chanfray and Marteau \cite{Martini:2009uj}, in the following called ``MECM model''. 
It is based on the nuclear response functions calculated in random phase approximation and allows an unified treatment of the quasielastic, 
the multinucleon emission channel and the coherent and incoherent pion production. The agreement with the experimental data 
in the pion production channels \cite{:2008eaa,AguilarArevalo:2009eb,Kurimoto:2009wq} has been proved.
Nevertheless the main feature of this MECM model 
is the treatment of the multinucleon emission channel in connection with the quasielastic. In fact,
as suggested in \cite{Martini:2009uj,Martini:2010ex}, the inclusion of this channel in the quasielastic cross section is 
a possible explanation of the MiniBooNE quasielastic total cross section \cite{AguilarArevalo:2010zc}, 
apparently too large with respect to many theoretical predictions \cite{AlvarezRuso:2010ia} 
employing the standard value of the axial mass. Since the MiniBooNE experiment, as well as many others involving Cherenkov detectors, defines 
a ``quasielastic'' event as the one in which 
only a final charged lepton is detected, the ejection of a single nucleon
(a genuine quasielastic event) is only one possibility, and one
must in addition consider events involving a correlated nucleon pair from
which the partner nucleon is also ejected. This leads to the excitation of
2 particle-2 hole (2p-2h) states; 3p-3h excitations are also possible. 
Nowadays other models 
\cite{Nieves:2011pp,Amaro:2011qb,Bodek:2011ps,Lalakulich:2012ac}
have included the multinucleon contribution in the 
computation of the cross sections relevant for the MiniBooNE quasielastic 
kinematics, improving the agreement with the experimental data. 
For a brief review see for example \cite{Martini:2011ui}. 
Recently, it has been shown \cite{Martini:2011wp} that the MECM model can also reproduce 
the MiniBooNE flux averaged double differential cross section \cite{AguilarArevalo:2010zc} which 
is a directly measured quantity and hence free from the model-dependent uncertainties in the neutrino energy 
reconstruction, and 
the total inclusive cross section \cite{Martini:2011ui} (also employed by T2K as described below) measured by SciBooNE \cite{Nakajima:2010fp}.
In the following we will use the cross sections obtained in the two different approaches described above in several exclusive channels (quasielastic and pion production), 
as well as in the inclusive one, for both charged current (CC) and neutral current (NC) interactions on carbon and oxygen (the targets used in near and far T2K detectors, respectively)
and for two neutrino flavours $\nu_\mu$ and $\nu_e$. Although all exclusive channels are involved in the analysis, 
we will refer to the first model as ``the Fermi Gas model''
and to the second approach as ``the MECM model''. 

\begin{figure}[!h]
\centering
\includegraphics[width=0.65\textwidth]{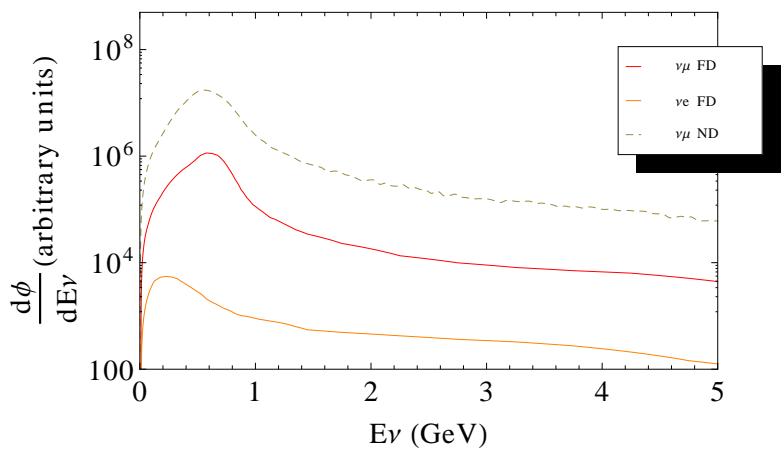}
\caption{\it Fluxes at near ($\nu_\mu$ only) and far ($\nu_\mu$ and $\nu_e$) detectors.}
\label{fluxes}
\end{figure}

\begin{figure}[!h]
\centering
\includegraphics[width=0.55\textwidth]{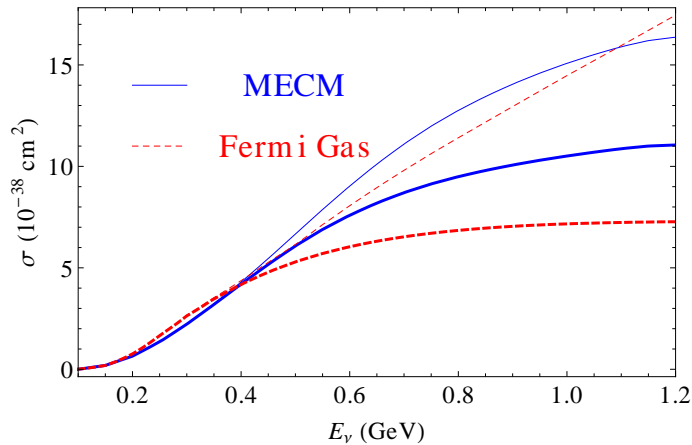}
\caption{\it Inclusive (thin lines) and QE (thick lines) $\nu_\mu$ CC cross sections on oxygen in the FG model (dashed lines) and in  the MECM model (solid line) 
after the normalization of the inclusive cross sections to the event rate at the ND. }
\label{xs}
\end{figure}

In order to perform our comparison among the above-mentioned models, we first need to correctly normalize the 
Fermi gas to the T2K event rates, at both near (ND) and far (FD) detectors; we
use the following algorithm:
\begin{itemize}
\item[{1-}] normalization of the cross section with the  $\nu_\mu$ inclusive CC at the ND; according to \cite{Abe:2011sj},
we have to reproduce 1529 $\nu_\mu$ inclusive events, collected using $2.9 \times 10^{19}$ POT, in the energy range $[0-5]$ GeV, with an active 
detector mass of 1529 Kg \footnote{We thank Scott Oser for providing such a number to us.} at a distance of 280 m from the $\nu$ source and half a year of data taking (Run 1). Notice that only the muon neutrino 
cross sections can be correctly normalized; we assume that the same normalization also applies for the $\nu_e$ cross section, although they could differ at the 
$\mu$ production threshold  (in  any case
away from the peak of the neutrino flux);
\item[{2-}] computation of the expected events (and energy distributions) at the far detector in the appropriate two-parameter plane ($(\sin^2 2\theta_{13},\delta_{CP})$ 
for appearance and $(\theta_{23},\Delta m^2_{atm})$ for disappearance); 
\item[{3-}] normalization to the T2K spectral distributions.  
\end{itemize}
Step \#3 is needed to get rid of the experimental efficiencies applied by the T2K collaboration to the signal and background events. This means that 
the bin contents of our simulated distributions (obtained at point \#2) are corrected by coefficients, generally of ${\cal O}(1)$ that we consider as a detector property,
and then not further modified.
For a different model, we repeat step \#1 and then go to step \#2, using the same 
normalization coefficients extracted in step \#3 with the Fermi gas.
We make use of the GloBES \cite{Huber:2004ka} and MonteCUBES \cite{Blennow:2009pk} softwares
for the computation of event rates (and related $\chi^2$ functions) expected at the T2K  ND and FD detectors.
The fluxes of $\nu_\mu$, $\nu_e$ and their CP-conjugate counterparts predicted at the FD in absence of oscillations have been extracted directly from Fig.1 of \cite{Abe:2011sj},
whereas the $\nu_\mu$ flux at the ND has been obtained from \cite{giganti}. Such fluxes (the relevant ones summarized in Fig.\ref{fluxes}) are given for $10^{21}$ POT.
\noindent
As already stressed, for the relevant cross sections  we assumed that the T2K collaboration uses some ``sophisticated'' version of the Fermi gas model \cite{Smith:1972xh}.
In Fig.\ref{xs} we show the inclusive and QE cross sections in the FG model (dashed lines) and in  the MECM model (solid line) used in 
our simulation, after having correctly normalized the inclusive cross sections to the event rate at the ND. Especially for the MECM model, this procedure involves 
a degree of extrapolation of the inclusive cross sections towards neutrino energies beyond the validity of model itself. However, neutrino fluxes 
above ${\cal O}(1)$ GeV drop very fast and we checked that different kind of extrapolations do not alter our conclusions.

The important feature here is that, even after the normalization procedure, the MECM CCQE cross section is still larger than the FG predictions,
in the energy range relevant for appearance studies. This is due to the inclusion of the multinucleon component and 
will be the main reason of the differences between the results obtained in the two models. 
Note on the contrary that the inclusive cross sections are not really different. 

\section{The appearance channel}
The $\nu_\mu \to \nu_e$ transition probability is particularly suitable for extracting information on $\theta_{13}$ and 
$\delta_{CP}$; at the T2K energies ($E_\nu$) and baseline (L), one can expand the full 3-flavour probability
up to second order in the small parameters $\tetaot, \Delta_{12} / \Delta_{13}$ and $\Delta_{12} L$, with $\Delta_{ij}=\Delta m^2_{ij}/4 E_\nu$ \cite{Cervera:2000kp}:

\bea
P_{\nu_\mu\to \nu_e} & = & 
s_{23}^2 \, \sin^2 2 \tetaot \, \sin^2 \left (\delot \, L \right ) + 
c_{23}^2 \, \sin^2 2 \theta_{12} \, \sin^2 \left( \Delta_{sol} \, L \right ) \nn \\
& + & \tilde J \, \cos \left (\delta_{CP} + \delot \, L \right ) \;
(\Delta_{sol} \, L)\, \sin \left ( 2\, \delot \, L \right ) \, , 
\label{vacexpand} 
\eea
where 
\be
\tilde J \equiv c_{13} \, \sin 2 \theta_{12} \sin 2 \tetatt \sin 2 \tetaot\,, \qquad s_{23} = \sin \theta_{23}\,.
\ee
We clearly see that CP violating effects are encoded in the interference term proportional to the product of the solar mass splitting and the baseline, 
implying a scarce dependence of this facility on $\delta_{CP}$ when only the $\nu_\mu\to \nu_e$ channel (and the current luminosity) is considered.

\subsection{Extracting the T2K data}
Events in the far detector (obtained with $2.9 \times 10^{20}$ POT) are $\nu_e$ CCQE from $\nu_\mu \to \nu_e$ oscillation, with main backgrounds given 
by $\nu_e$ contamination in the beam and neutral current events with a misidentified $\pi^0$.  
The experimental data have been grouped in 5 reconstructed-energy bins, from 0 to 1.25 GeV and they are summarized in Tab.\ref{exp}.
The expectations for signal and backgrounds have been computed by the T2K collaboration from Monte Carlo simulations, for fixed value 
of the oscillation parameters, namely $\sin^2 2\theta_{12}=0.8794,\sin^2 2\theta_{13}=0.1, \sin^2 2\theta_{23}=1$ and 
$\Delta m^2_{sol}=7.5\times 10^{-5} eV^2,\Delta m^2_{atm}=+2.4\times 10^{-3} eV^2 $. In order to normalize our event rates to
the T2K Monte Carlo expectations, we extracted these numbers from Fig.5 of \cite{Abe:2011sj} and reported them in Tab.\ref{exp}.
\begin{table}[h!]
\begin{center}
\begin{tabular}{c||c|c|c|c|c|c||c}
\hline \hline
         & channel        & bin 1  &   bin2 & bin3 & bin4 & bin5 & total  \\
\hline
{\it exp data} & & 0 & 2  & 2 & 1 & 1& 6\\
\hline
{\it estimates}    & $\nu_\mu \to \nu_e$   & 0.197  &  0.991   & 2.008  & 0.783  & 0.192  &  4.171 \\
{\it for} & $\nu_e \to \nu_e$    & 0.025  &  0.162   & 0.204  & 0.158  & 0.113  & 0.662       \\
{\it $\sin^2 2\theta_{13}=0.1$}& NC    & 0.07 &  0.227   & 0.148  & 0.08  & 0.04  & 0.565 \\ \hline
\end{tabular}
\caption{\it \label{exp} Expected event rates for $\sin^22\theta_{13}$=0.1.}
\end{center}
\end{table}
 
Notice that we used the central bin energy as a reference value for the neutrino energy in a given bin; this could be 
different from the reconstructed neutrino energies used by the T2K collaboration. To mimic 
possible uncertainties associated to the neutrino energy reconstruction, 
we apply an energy smearing function to distribute the rates in the various energy bins. 
Other choiches, more related to microscopical calculations 
\cite{Benhar:2009wi,Leitner:2010kp,Martini:2012fa} 
are also possible. In particular, an analysis of the validity of the approximation
contained in the identification of the reconstructed neutrino energy
via a two-body kinematics with the real neutrino energy has been done in 
\cite{Martini:2012fa}, where the MECM model has been employed. The role of several nuclear
effects such as Pauli blocking, Fermi motion, RPA correlations and
multinucleon component has been studied in details. This analysis was
performed, among others, considering the T2K conditions at near and
far detectors, paying a particular attention to the $\nu_\mu \to
\nu_e$ appearance mode.
The ratios among our computation and the T2K data ({\it energy dependent} efficiencies) are 
summarized in Tab.\ref{tabella}.
This procedure (corresponding to step \#3 of the previous paragraph) allows us to take into account all the detection efficiencies to different neutrino flavours
in the Super Kamiokande detector. Once computed, these corrective factors are used in the simulations done with a different cross section, since we assume here that they are features
of the detector and not of the neutrino interactions. 
\begin{table}[h!]
\begin{center}
\begin{tabular}{cccccc}
\hline \hline
channel  &  bin 1  & bin 2  &bin 3  &bin 4  &bin 5    \\
\hline
 $\nu_\mu \to \nu_e$  &  1.76&1.42&1.52&1.72&1.90     \\
 $\nu_e \to \nu_e$ & 1.10&1.60&1.65&1.55&1.70   \\
NC &  0.04&0.025&0.009&0.01&0.016   \\ \hline
\end{tabular}
\caption{\it \label{tabella} Efficiencies computed after normalizing the event rates at the values for $\sin^22\theta_{13}$=0.1.}
\end{center}
\end{table}
We see that for $\nu_{e,\mu} \to \nu_e$ transitions these numbers are just ${\cal O}(1)$ coefficients, which makes us 
confident that the normalization procedure correctly accounts for the main experimental features. The same 
is not true for the  NC events which, however, have not been normalized to the ND as for the CC interactions.
As a check, we also computed the expected events for $\sin^22\theta_{13}$=0, obtaining 0.1 
$\nu_\mu \to \nu_e$ events and 0.72  $\nu_e \to \nu_e$ events (and the same neutral current rate), in good agreement with 
the T2K expectations \cite{Abe:2011sj}.
 
\subsection{Fit to the data}
\begin{figure}[!h]
\centering
\includegraphics[width=0.55\textwidth]{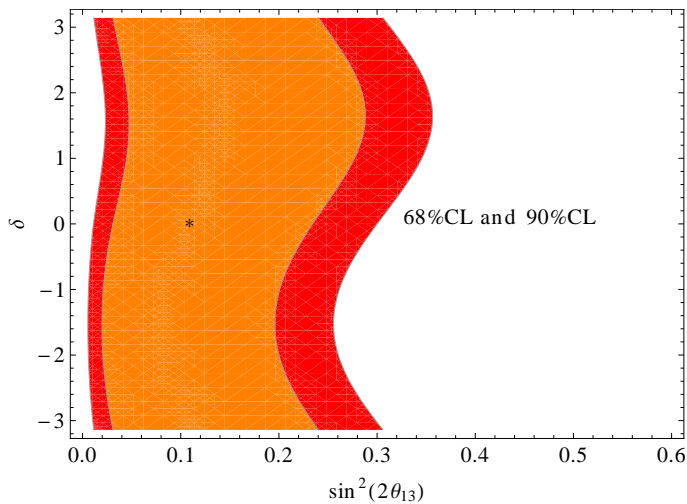}
\caption{\it The 68\% and 90\%~C.L. regions for $(\sin^{2}2\theta_{13},\delta_{\rm CP})$ in the FG model.}
\label{fit}
\end{figure}

Equipped with these results, we performed a $\chi^2$ analysis to reproduce the confidence level regions in the 
$(\sin^2 2\theta_{13},\delta_{CP})$-plane shown in Fig.6 of  
\cite{Abe:2011sj}. Contrary to what has been done in the official T2K paper, we make a complete three-neutrino 
analysis of the experimental data, marginalizing over all parameters not shown in the confidence regions. As external
input errors, we used 3\% on $\theta_{12}$ and  $\Delta m^2_{sol}$, 8\% on $\theta_{23}$ and  6\% on $\Delta m^2_{atm}$.
We use a constant energy resolution function $\sigma(E_\nu)=0.085$ and, for simplicity, we adopt a 7\% normalization error
for the  signal and 30\% for the backgrounds. We also used energy calibration errors fixed to $10^{-4}$ for the signal and $5\cdot 10^{-2}$ for the backgrounds;
normalization and energy calibration errors take into account the impact of systematic errors in the $\chi^2$ computation.\\
Assuming a normal hierarchy spectrum, the best-fit point from the fit procedure is (obviously):
\bea
\sin^2(2\theta_{13})=0.108 \qquad \delta_{CP} = 0.04
\eea 
with $\chi^2_{min} = 1.69$; the related contour plot is shown in Fig.\ref{fit}.
Compared to the official release, the plot is in quite good agreement, although the allowed values of $\theta_{13}$ around 
maximal CP violation $\delta_{CP}=\pi/2$ are a bit larger (this is the effect of including the errors of 
the atmospheric parameters into the fit procedure).

We now apply the same procedure to determine $\theta_{13}$ using the MECM cross sections described in \cite{Martini:2009uj}. 
In doing that, we normalize the 
cross sections to the ND events and then compute the number of oscillated events (and related backgrounds), to be compared 
with the experimental T2K data. We assume that the efficiencies reported in Tab.\ref{tabella} are exactly the same, since they are a property 
of the SK detectors and then independent on the cross section model. With these assumptions, we get the following number of expected rates
for $\sin^22\theta_{13}$=0.1:

\begin{table}[h!]
\begin{center}
\begin{tabular}{c||c|c|c|c|c|c||c}
\hline \hline
         & channel        & bin 1  &   bin2 & bin3 & bin4 & bin5 & total  \\
\hline
{\it estimates}    & $\nu_\mu \to \nu_e$   & 0.234  &  1.205   & 2.808  & 1.121 & 0.295  &  5.665 \\
{\it for} & $\nu_e \to \nu_e$    & 0.029  &  0.194   & 0.280  & 0.227  & 0.179  & 0.909      \\
{\it $\sin^2 2\theta_{13}=0.1$}& NC    & 0.017 &  0.156   & 0.204  & 0.130  & 0.08  & 0.590 \\ \hline
\end{tabular}
\caption{\it \label{tabellaM} Total rates for $\sin^22\theta_{13}$=0.1 in the MECM model.}
\end{center}
\end{table}

\begin{figure}[!h]
\centering
\includegraphics[width=0.55\textwidth]{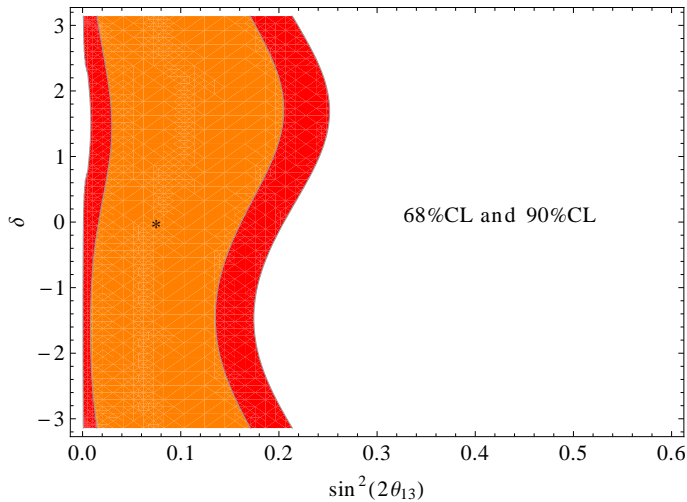}
\caption{\it The 68\% and 90\%~C.L. regions for $(\sin^{2}2\theta_{13},\delta_{\rm CP})$ for the MECM model. Star indicates 
the best fit point.}
\label{fitM}
\end{figure}

\begin{figure}[!h]
\centering
\includegraphics[width=0.55\textwidth]{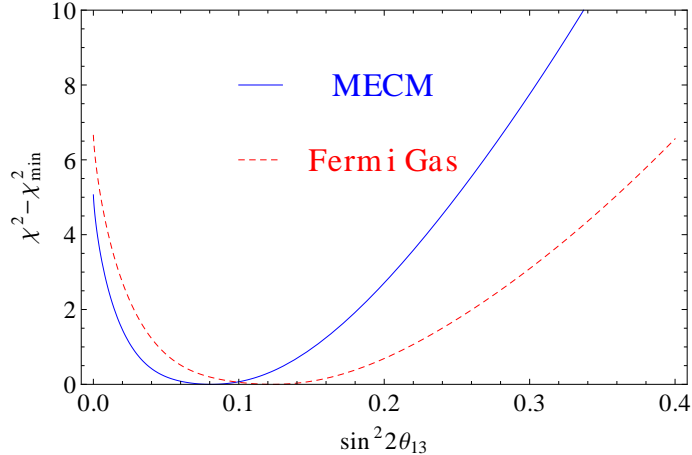}
\caption{\it $\chi^2$ as a function of $\theta_{13}$ for the MECM model (solid line) and FG (dashed line).}
\label{chi2}
\end{figure}

It is clear that larger rates need smaller $\theta_{13}$ to reproduce the data (the effect of the CP phase $\delta$ is negligible with 
such a statistics). The best fit point is:
\bea
\sin^2(2\theta_{13})=0.073 \qquad \delta_{CP} = 0\,,
\eea 
with $\chi^2_{min} = 1.53$, and the contour plot is shown in Fig.\ref{fitM}.
We can appreciate a substantial improvement in the determination of the reactor angle, whose largest value is 0.24, to be compared with 0.35 obtained with the Fermi Gas.
To make a more direct comparison on $\theta_{13}$ between the FG and MECM results, in Fig.\ref{chi2} we show the $\chi^2-\chi^2_{min}$ function, computed marginalizing over all 
other oscillation parameters (including $\delta_{CP}$). At 1$\sigma$, we get:
\bea
\sin^2 2\theta_{13}^{MECM} &=& 0.08^{\left(^{+0.07}_{-0.05}\right)}
\nn \\ \label{1sigma} \\
\sin^2 2\theta_{13}^{FG} &=& 0.12^{\left(^{+0.08}_{-0.09}\right)} \nn\,.
\eea
They are clearly compatible although, as expected, $\theta_{13}^{MECM}  < \theta_{13}^{FG} $.

\section{The disappearance channel}
\begin{figure}[!h]
\centering
\includegraphics[width=0.45\textwidth]{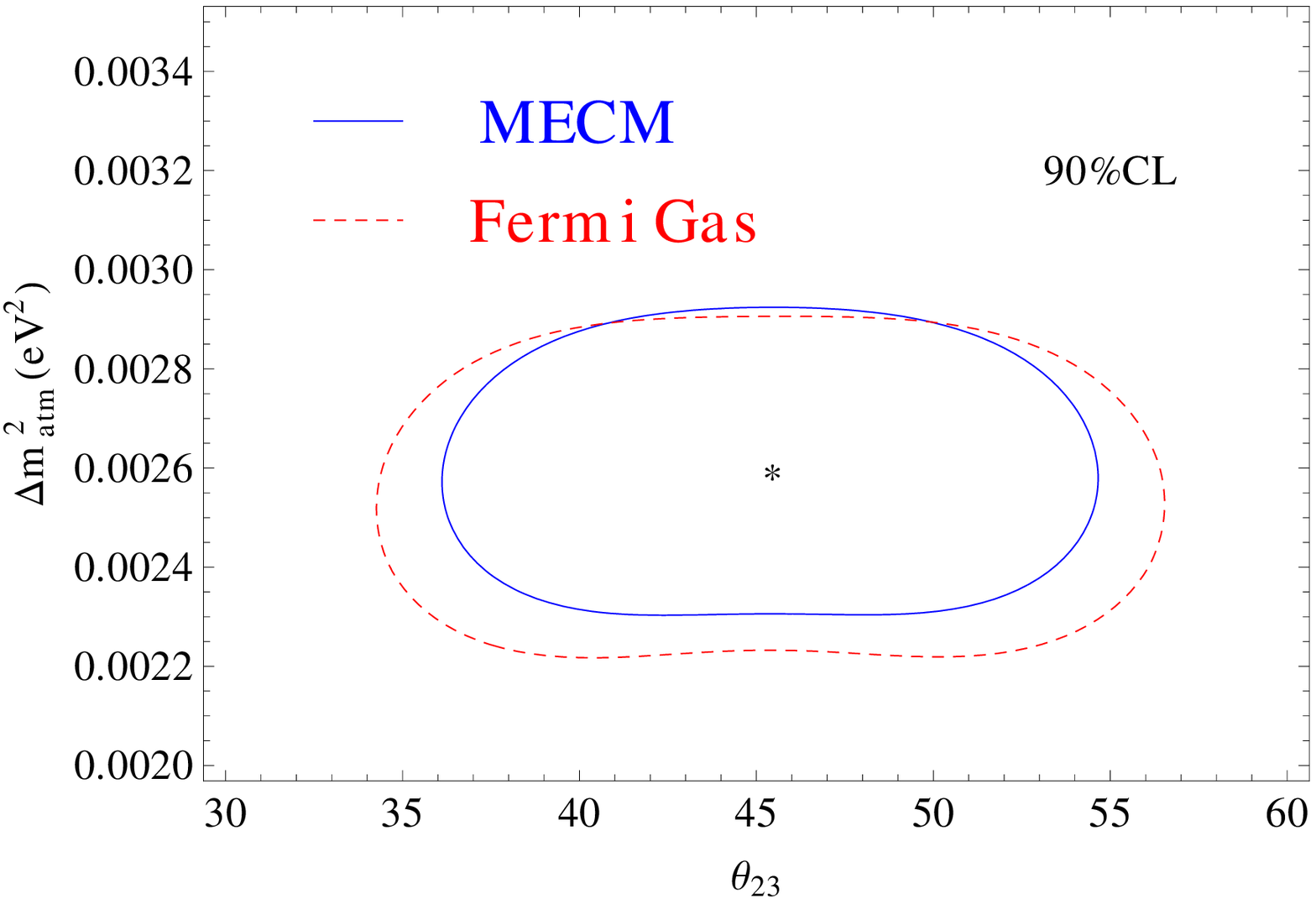}
\includegraphics[width=0.45\textwidth]{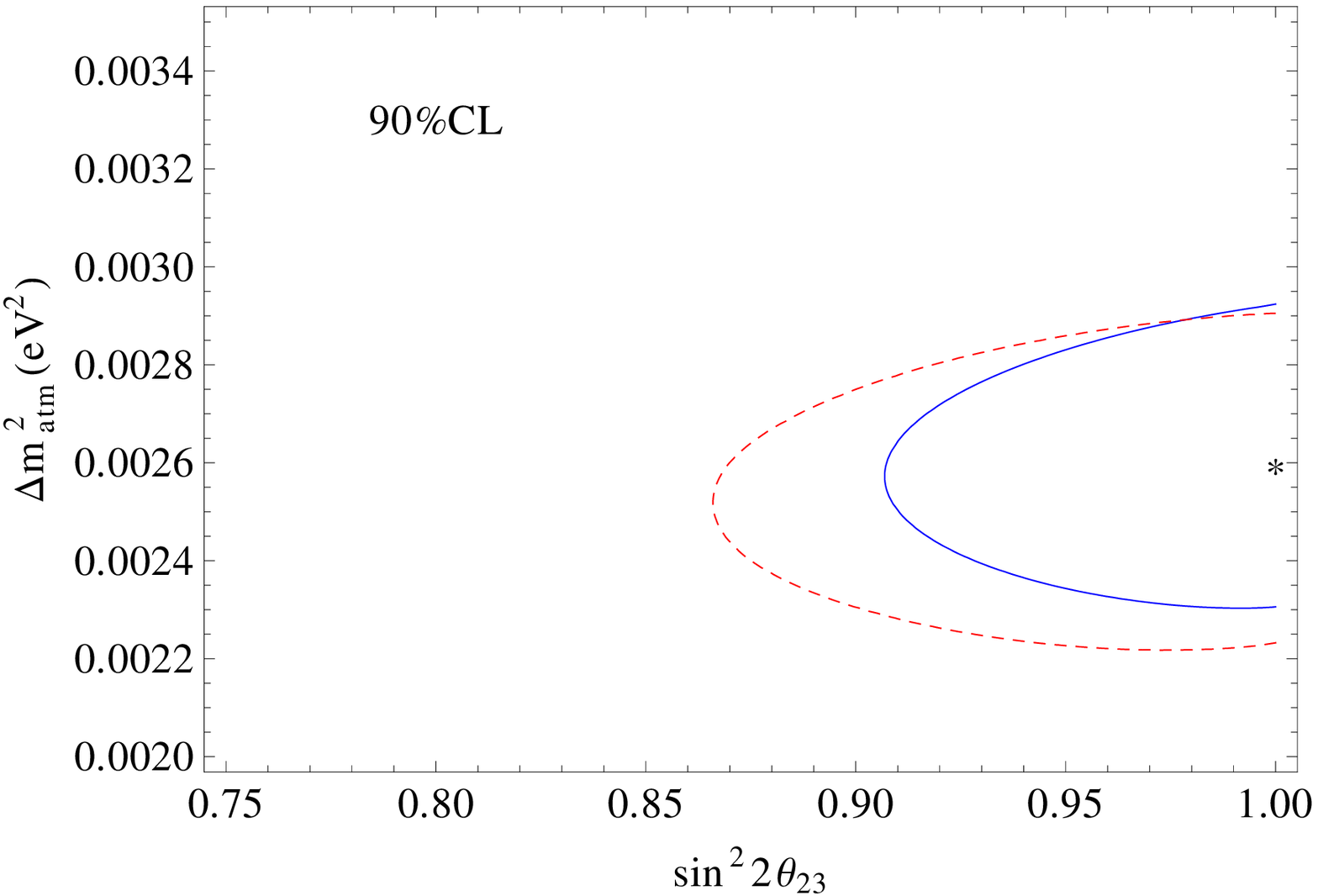}
\caption{\it 90\% contour levels for the MECM model (solid line) and FG (dashed line), 
in the $(\theta_{23},\Delta m^2_{atm})$ (left panel) and 
 $(\sin^2 2\theta_{23},\Delta m^2_{atm})$ (right panel) planes. Star indicates the best fit obtained in the  MECM model.}
\label{distot}
\end{figure}

We extend the previous analysis to include the first 
disappearance $\nu_\mu \to \nu_\mu$ data \cite{giganti}. 
In the two-flavour limit, (the one where both $\theta_{13}$ and  $\Delta m^2_{sol}$ are vanishing)
the $\nu_\mu \to \nu_\mu$ probability reads \cite{Donini:2005db}:
\bea
P(\nu_\mu \to \nu_\mu) & = & 1-   \sin^2 2 \theta_{23} \, \sin^2 \left(\Delta_{atm} L\right )
 \label{eq:probdismu}\,.
\eea
Effects related to $\theta_{13}$ are clearly sub-dominant, so that this channel is particularly useful to extract 
information on the atmospheric parameters. 
The T2K collaboration collected 31 data events,
grouped in 13 energy bins, as one can see from Fig.3 of \cite{giganti}. The sample extend up to
6 GeV and it is mainly given by $\nu_\mu$CCQE, $\nu_\mu$CC non-QE, $\nu_e$ CC and NC.
Differently from the appearance channel, we cannot normalize their energy distribution to the channel-by-channel T2K Monte Carlo 
expectation since, as far as we know, such information has not been released. We can only normalize our FG cross 
section to the total rates shown in Tab.I of \cite{giganti}, which amounts to 17.3, 9.2, 1.8 and $<$0.1 
events for $\nu_\mu$CCQE, $\nu_\mu$CC non-QE, NC and $\nu_e$ CC, respectively. Such numbers
refer to $\sin^2(2\theta_{23})=1.0$ and $|\Delta m^2_{atm}|=2.4 \times 10^{-3}$ eV$^2$, with all other
neutrino mixing parameters vanishing.
For the sake of completeness, we summarize in  Tab.\ref{tab2} the T2K data as well as the energy distributions of the 
$\nu_\mu$CCQE and $\nu_\mu$CC non-QE as obtained using the MECM cross sections.
In evaluating such numbers, we assume a variable bin size, centered in the neutrino energy corresponding to the T2K data.
In our fit procedure we have assumed a total normalization for the NC as given in \cite{giganti}, but with
appropriate energy distributions  according to the FG and MECM cross sections. 
We have also adopted a conservative 15\% normalization error and energy calibration error at the level of $10^{-3}$  
for both signal and background.
\begin{table}[h!]
\begin{center}
\begin{tabular}{cccccccccccccc}
\hline \hline
 bin  &    1  &  2  & 3  & 4  & 5  &   6  &  7  & 8  & 9  & 10 &   11  &  12  & 13   \\
\hline
 {\it T2K data} & 1 & 5 & 3 & 1 & 2 & 1 & 2 & 3 & 4 & 2 & 1 & 3&  3  \\
{\it MECM} $\nu_\mu$CCQE &  0.6 & 3.2&  2.2 & 0.7&  1.8& 0.8&
2.0& 2.8& 3.5& 1.2& 1.3& 0.8& 0.6
\\ 
{\it MECM} $\nu_\mu$ CC non-QE &  0.0 & 0.0&  0.0 & 0.0&  0.0& 0.3&
0.2& 0.4& 0.3& 0.4& 1.0& 1.2& 1.3\\
\hline 
\end{tabular}
\caption{\it \label{tab2} T2K events and bin distributions for the $\nu_\mu$CCQE and $\nu_\mu$ CC non-QE rates in the  MECM model.}
\end{center}
\end{table}
The results of our fit procedure can be appreciated in Fig.\ref{distot}, where we show the 90\% CL for the Fermi Gas (dashed line) and the MECM 
model (solid line), in the case of normal hierarchy. We plot the 2 degrees of freedom (dof) confidence levels in the $(\theta_{23},\Delta m^2_{atm})$ (left panel) and 
 $(\sin^2 2\theta_{23},\Delta m^2_{atm})$ (right panel, to facilitate the comparison with the official T2K results) planes.
Again, the plots have been obtained marginalizing over the not shown parameters (a full three-flavour analysis); we considered a 50\% error on $\sin ^2 2\theta_{13}$ 
(with best fit at $\sin ^2 2\theta_{13}=0.0059$) and $\delta_{CP}$ undetermined.
We obtained:
\bea
&\text{FG}:& \sin^2 2\theta_{23} > 0.86 \qquad    2.22 \cdot 10^{-3}<\Delta m^2_{atm} \,(\text{eV}^2)< 2.90 \cdot 10^{-3} \nn  \\
&& \\
&\text{MECM}:& \sin^2 2\theta_{23} > 0.91 \qquad    2.31 \cdot 10^{-3}<\Delta m^2_{atm}\, (\text{eV}^2)< 2.93 \cdot 10^{-3} \nn 
\eea
with best fit points:
\bea
&\text{FG}:& \sin^2 2\theta_{23} =0.99 \; (47.9^\circ) \qquad    \Delta m^2_{atm} = 2.56 \cdot 10^{-3} \,\text{eV}^2\nn  \\
&& \\
&\text{MECM}:& \sin^2 2\theta_{23} =1.00 \; (45.0^\circ) \qquad    \Delta m^2_{atm} = 2.62 \cdot 10^{-3}\, \text{eV}^2\nn \,.
\eea
Some comments are in order;
first of all, we observe that, for both models, the best fit point is different from the T2K ones, which is 
\bea
(\sin^2 2\theta_{23})_{T2K} =0.98  \qquad    |\Delta m^2_{atm}|_{T2K} = 2.65 \cdot 10^{-3} \,\text{eV}^2\nn\,;
\eea
this is somehow obvious since we normalized our events to the MC predictions obtained for a different set of atmospheric 
parameters. The MECM cross section gives a better determination of both $\theta_{23}$ and $\Delta m^2_{atm}$,
mainly due to the larger statistics than the FG; at the same time, the disappearance
probability in Eq.(\ref{eq:probdismu}), for negligible solar mass difference and reactor angle, is smaller if the 
atmospheric mass difference is larger, for fixed  $\sin^2 2\theta_{23}$. This is what happens here, where a smaller
$P(\nu_\mu \to \nu_\mu)$ (and then a larger  $\Delta m^2_{atm}$) is needed in the MECM model to partially compensate for 
the larger cross section.

\section{Future perspectives}
\begin{figure}[!h]
\centering
\includegraphics[width=0.45\textwidth]{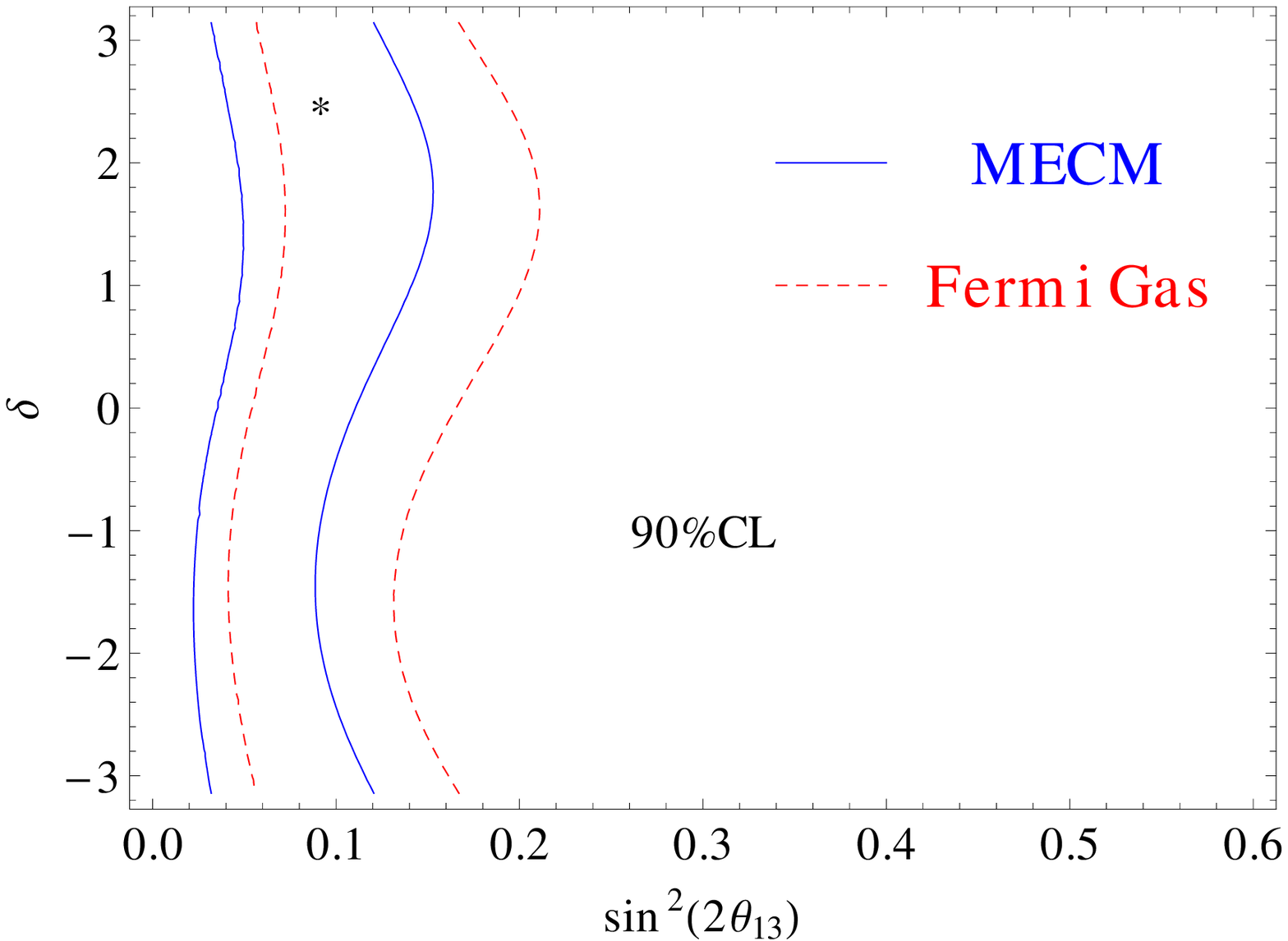}
\includegraphics[width=0.475\textwidth]{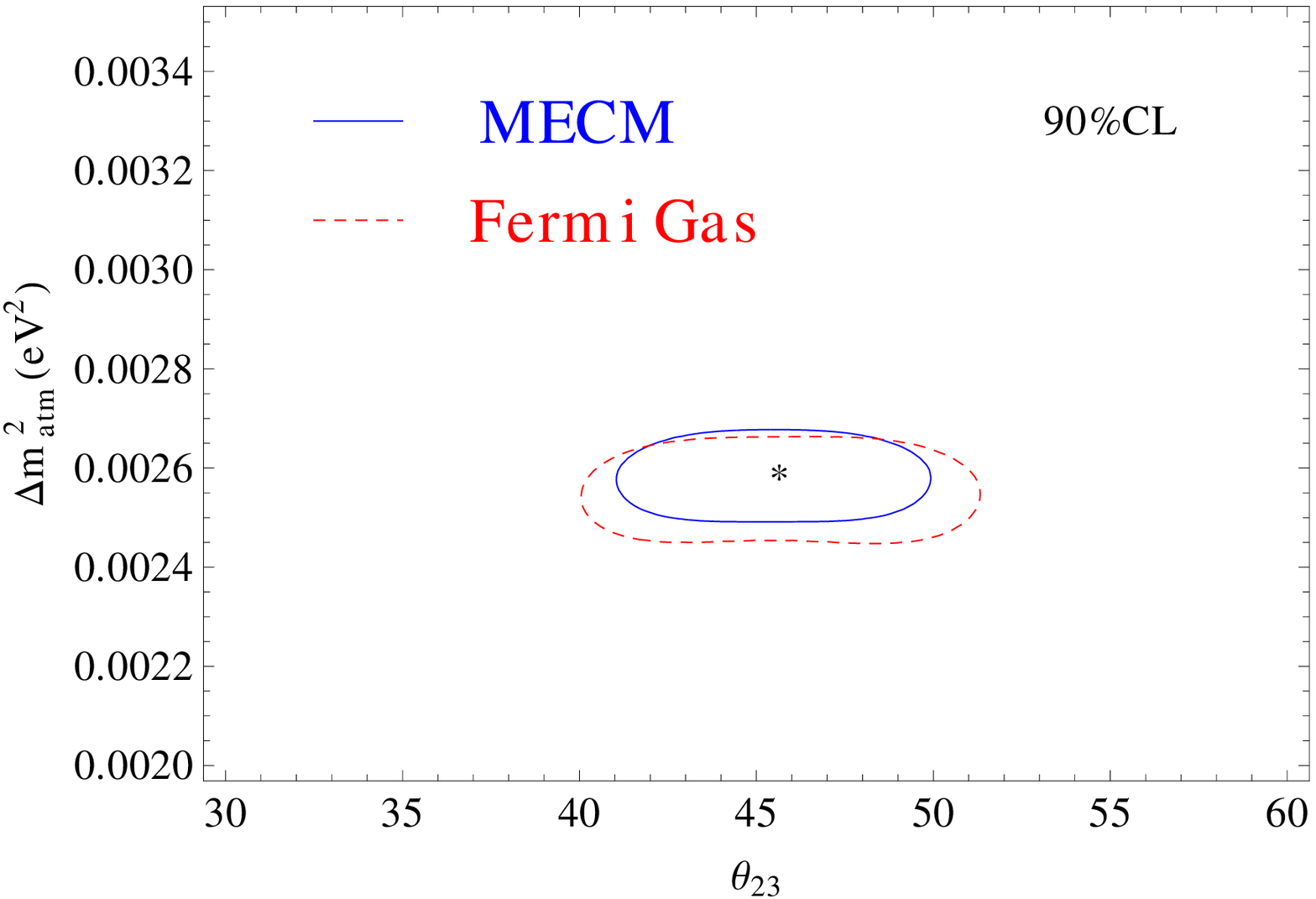}
\caption{\it 90\% CL for 2 dof in the $(\sin^2 2\theta_{13},\delta_{CP})$-plane (left panel) and 
$(\theta_{23},\Delta m^2_{atm})$-plane (right panel)  for the MECM model (solid line) and FG (dashed one) in the case the current T2K statistics 
is increased by a factor of 10. Stars indicate the best fit values of the parameters as obtained in the MECM model.}
\label{fin2dof}
\end{figure}

The statistics used by the T2K collaboration to make the disappearance study is only a 2\% of the rates expected at the end
of the experiment. It makes sense to ask how the previous results would modify if the accumulated statistics 
would be larger than the current one. We limit ourselves to consider a  number of events with the same energy distribution
as the experimental ones but bin contents larger by factor of 10, in both 
appearance and disappearance channels. In the analysis of the appearance channel, 
the (weak) information on $\theta_{13}$ contained in the disappearance sample should not be neglected (as we did previously); at the 
same time,  the dependence on the atmospheric parameters from the appearance sample cannot in principle be neglected when studying the 
disappearance data. For this reason, we prefer to combine both $\nu_\mu \to \nu_e$ and $\nu_\mu \to \nu_\mu$ oscillation data, and study the sensitivity 
to the reactor and atmospheric parameters as we did in the previous sections, marginalizing over the parameters not expressly shown.
Notice that such an approach would not give any additional information on the mixing parameters if adopted with the current T2K statistics:
in fact, we see from Fig.\ref{distot} that the uncertainties on $\theta_{23}$ and $\Delta m^2_{atm}$ obtained from the T2K data
are larger than the adopted external errors on these parameters in the appearance channel, so that 
adding the $\nu_\mu \to \nu_\mu$ data will not improve the sensitivity to $\theta_{13}$; similarly, the dependence on the reactor 
angle in $P(\nu_\mu \to \nu_\mu)$ is sub-leading and the impact of the disappearance channel in the appearance measurement is also negligible.
We stress that extracting information on the mixing parameters combining appearance and disappearance channels
is also mandatory to avoid some inconsistencies emerged in the official T2K fits,
where $|\Delta m^2_{atm}|$ is fixed to $2.4 \times 10^{-3}$ eV$^2$  in the appearance analysis whereas the 
best fit point obtained from the disappearance data is  $2.6 \times 10^{-3}$ eV$^2$.
The results of our procedure are shown in Fig.\ref{fin2dof}, where we display the 90\% CL in the $(\sin^2 2\theta_{13},\delta_{CP})$-plane (left panel) and 
$(\theta_{23},\Delta m^2_{atm})$-plane (right panel)  for the MECM (solid line) and FG (dashed one) models in the case the current T2K statistics 
is increased by a factor of 10.  The minimum of the $\chi^2$ still gets reasonable values: 
we obtain $\chi^2_{min}\sim$ 20 in the appearance analysis and $\chi^2_{min}\sim$ 30 in disappearance.
In both panels we can appreciate the effects of the increased statistics, as expected: for a given model of cross section, the allowed regions 
are strongly restricted with respect to the current situation. The best fit values for $\delta_{CP}$ are somehow different 
in the two models ($\delta_{CP}\sim 0$ and $\delta_{CP}\sim 144^\circ$ for the FG and MECM models, respectively), although
statistically non very significant.
Such intervals for $\sin^2 2\theta_{13}, \theta_{23}$ and 
$\Delta m^2_{atm}$ are summarized in Tab.\ref{finalX10} ($\delta_{CP}$ is obviously still unconstrained).
\begin{table}[h!]
\begin{center}
\begin{tabular}{cccc}
\hline \hline
  & $\sin^2 2\theta_{13}$ &  $\theta_{23}(^o)$ & $\Delta m^2_{atm} (10^{-3} \,\text{eV}^2)$\\
\hline
 FG  & [0.041-0.211] (0.105) & [40.1-51.3] (47.6) &  [2.45-2.67] (2.56)\\
 MECM & [0.023-0.154] (0.092) & [41.1-49.9] (45.4)& [2.49-2.67] (2.60)\\
 \hline
\end{tabular}
\caption{\it \label{finalX10} 90\% intervals for $\sin^2 2\theta_{13}, \theta_{23}$ and 
$\Delta m^2_{atm}$,  for the MECM  and FG models in the case the current T2K statistics 
is increased by a factor of 10. In parenthesis, the best fit points.}    
\end{center}
\end{table}
We have checked that, if we only use the appearance channel to extract $\theta_{13}$, the results are sligthly different: 
although the best fit value is practically 
indistinguishable from the one quoted in Tab.\ref{finalX10},
the confidence regions are a bit larger, with significant overlap with the above mentioned analysis. 
To see stronger effects due to the $\theta_{13}$ dependence in the $\nu_\mu \to \nu_\mu$ transition, we need a more accurate spectral information \cite{donini2}.
Similar conclusions can also be drawn for the disappearance channel: with only a factor of 10 more statistics and 
no appearance contribution, the best fit for the atmospheric parameters remain almost the same whereas the 
90\% CL region for $\theta_{23}$ shows a smaller lower limit (from $40.1^\circ$ to $39.8^\circ$) in the FG model.

\begin{figure}[!h]
\centering
\includegraphics[width=0.55\textwidth]{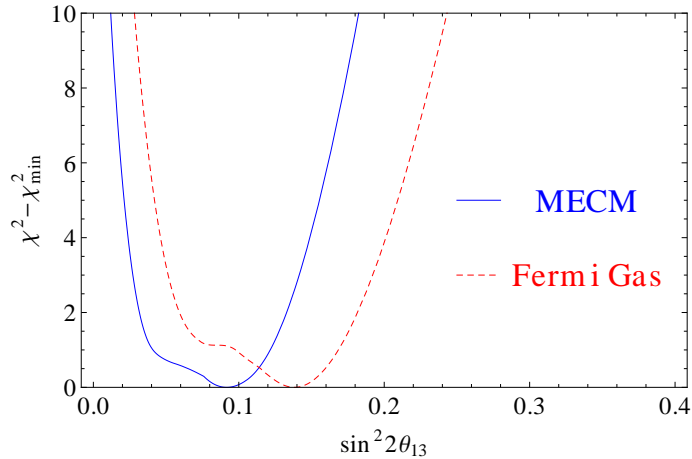}
\caption{\it $\chi^2-\chi^2_{min}$ as a function of $\sin^2 2\theta_{13}$ for the MECM model (solid line) and FG (dashed line) in the case the event rates 
are increased by a factor of 10. }
\label{chi2X10}
\end{figure}

Finally, we observe that such an increased statistics is necessary to make marginally incompatible the FG and MECM $\sin^2 2\theta_{13}$ results, see Fig.\ref{chi2X10}, obtained
marginalizing over $\delta_{CP}$ also.
In fact, at 1$\sigma$ we get:
\bea
\sin^2 2\theta_{13}^{MECM} &=& 0.092^{\left(^{+0.030}_{-0.052}\right)}
\nn \\ \label{1sigmax10} \\
\sin^2 2\theta_{13}^{FG} &=& 0.138^{\left(^{+0.031}_{-0.041}\right)} \nn\,.
\eea

\section{The inverted hierarchy case}
\begin{figure}[!h]
\centering
\includegraphics[width=0.45\textwidth]{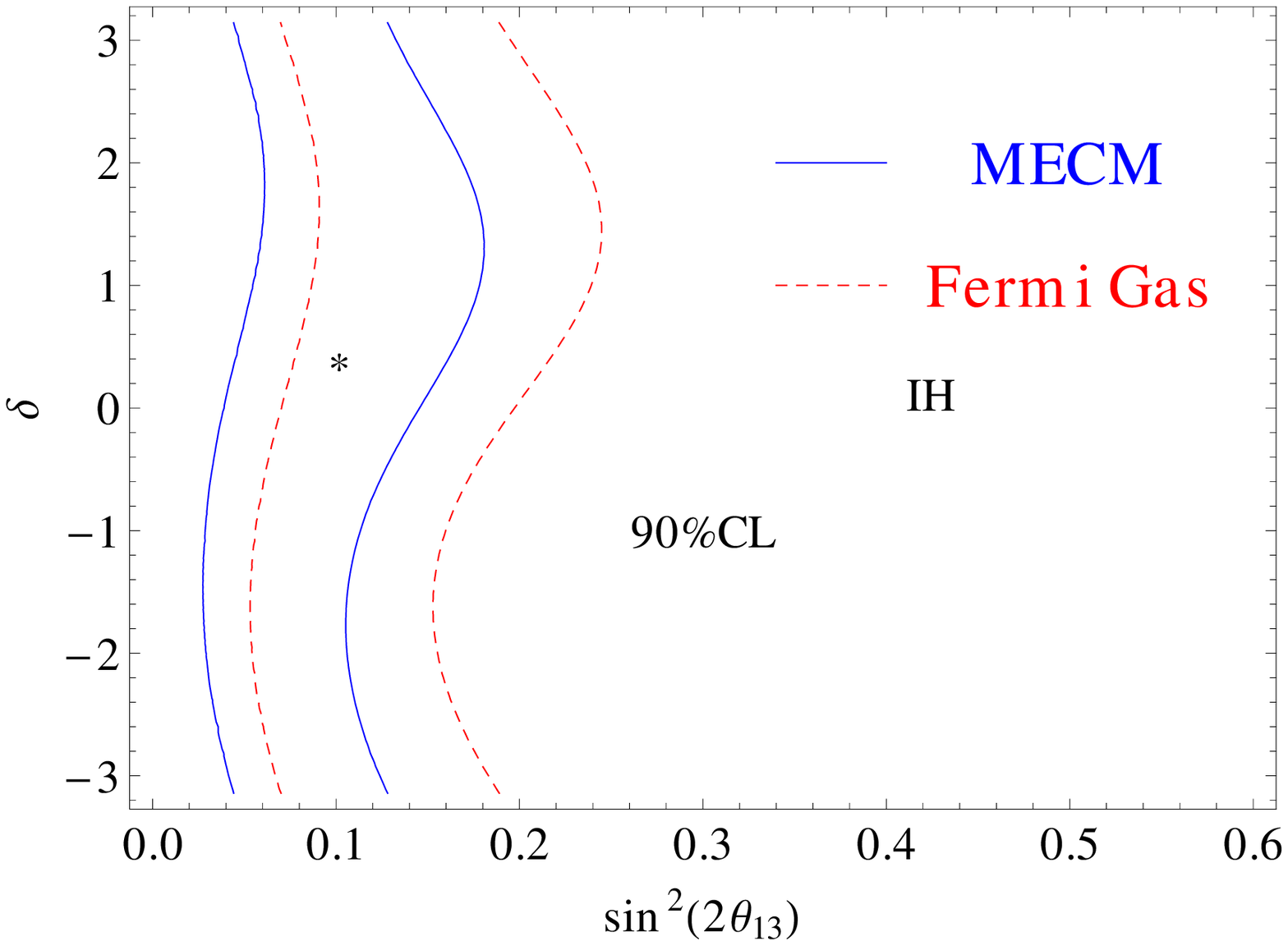}
\includegraphics[width=0.4825\textwidth]{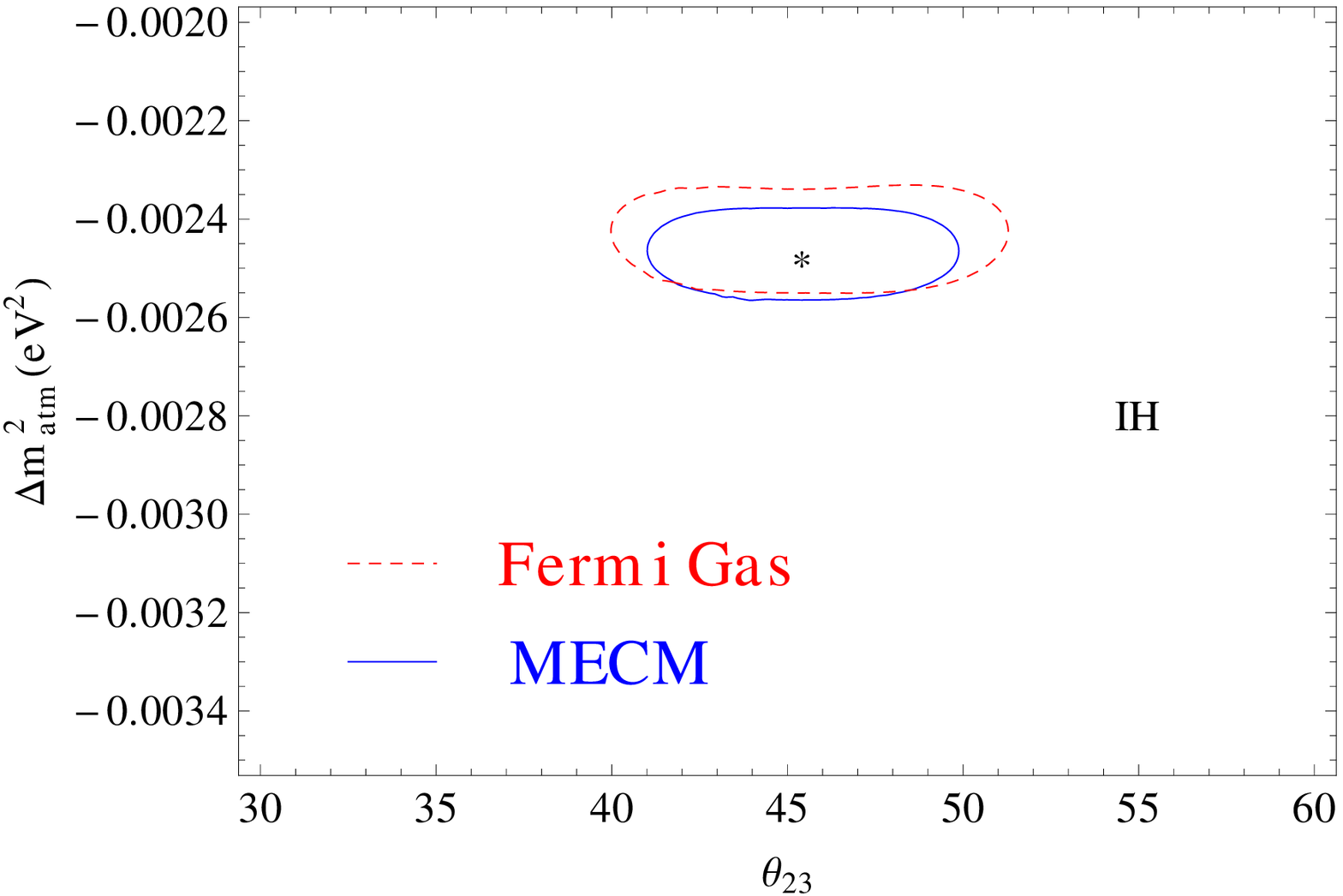}
\caption{\it 90\% CL for 2 dof in the $(\sin^2 2\theta_{13},\delta_{CP})$-plane (left panel) and 
$(\theta_{23},\Delta m^2_{atm})$-plane (right panel)  for the MECM model (solid line) and FG (dashed one) in the case the current T2K statistics 
is increased by a factor of 10 and the ``data'' are fitted with the inverted hierarchy. Stars indicate the best fit values of the parameters as obtained in the MECM model.}
\label{fin2dofIH}
\end{figure}

For the sake of completeness, we have repeated the same computations as above under the hypothesis that the neutrino mass spectrum
is of inverted type (IH). With the current T2K statistics, we cannot appreciate huge differences in the results obtained 
using the two different models for the cross section. Then, we limit ourselves here to the case where the stastistics is larger
by a factor of 10. Our results are summarized in Fig.\ref{fin2dofIH} and Tab.\ref{finalX10IH}.
\begin{table}[h!]
\begin{center}

\begin{tabular}{cccc}
\hline 
\hline
  & $\sin^2 2\theta_{13}$ &  $\theta_{23}(^o)$ & $|\Delta m^2_{atm}| (10^{-3} \,\text{eV}^2)$\\
\hline
 FG  & [0.049-0.241] (0.164) & [40.0-51.3] (47.6) &  [2.34-2.55] (2.44)\\
 MECM & [0.026-0.181] (0.102) & [41.1-49.8] (45.4)& [2.37-2.56] (2.47)\\
 \hline
\end{tabular}
\caption{\it \label{finalX10IH} 90\% intervals for $\sin^2 2\theta_{13}, \theta_{23}$ and 
$\Delta m^2_{atm}$,  for the MECM  and FG models in the case the current T2K statistics 
is increased by a factor of 10 and the ``data'' are fitted with the inverted hierarchy. In parenthesis, the best fit points.}    
\end{center}
\end{table}
Comparing the left panel of  Fig.\ref{fin2dofIH} with the corresponding one in Fig.\ref{fin2dof}, we clearly see that an inverted spectrum 
prefers larger values for $\theta_{13}$, in both models. The best fit of the CP phases is different among the two mass
orderings but not really significant. In the atmospheric plane, right panel of  Fig.\ref{fin2dofIH}, we observe the same pattern
as in the normal hierarchy case, that is the MECM tends to give a better resolution for both $\Delta m^2_{atm}$ and 
$\theta_{23}$ than the FG model.

\section{Conclusions}
In this paper we have studied the impact of using different models for the neutrino-nucleus cross section 
in the determination of the $\theta_{13,23}$ mixing angles and the atmospheric mass difference 
$\Delta m^2_{atm}$ using the recent T2K data, for both appearance and disappearance channels. 
Although the statistics is not large enough to draw definite conclusions, we have seen that a 
more refined treatments of nuclear effects in neutrino interactions can have some impact in the 
achievable precision on the mixing parameters. In particular, the MECM model predicts 
a large CCQE cross section, compared to the FG model, which results in a small $\theta_{13}$ needed to fit the data
in the $\nu_\mu \to \nu_e$ channel. At the same time, a larger $\Delta m^2_{atm}$ is required to fit the 
data in the $\nu_\mu$ disappearance channel, since a smaller disappearance probability is needed to compensate for the larger 
cross sections. Interestingly enough, with 10 times more statistics the two models tend to give substantial different 
results in terms of best fit points and parameter uncertainties (of course, better than before) but their predictions
are still compatible to each other.

\section*{Acknowledgements}
We are strongly indebted with Claudio Giganti for clarifying several technical aspects of the T2K experiment.
We also want to thank Enrique Fernandez-Martinez for useful suggestions about the fit procedure,
Lucio Ludovici for useful discussion on the flux normalization at the near detector in T2K.
D.M. acknowledges MIUR (Italy) for financial support under the program "Futuro in Ricerca 2010 (RBFR10O36O)". 
M.M. acknowledges the Communaut\'e
fran\c caise de Belgique (Actions de Recherche Concert\'ees) for financial support.


\begin{thebibliography}{99}

\bibitem{Abe:2011sj}
K.~Abe {\it et al.} [ T2K Collaboration ],
Phys.\ Rev.\ Lett.\  {\bf 107}, 041801 (2011).
[arXiv:1106.2822 [hep-ex]].


\bibitem{giganti}
K.~Abe {\it et al.}  [T2K Collaboration],
Phys.\ Rev.\ D {\bf 85}, 031103 (2012)
[arXiv:1201.1386 [hep-ex]].

\bibitem{FernandezMartinez:2010dm} 
  E.~Fernandez-Martinez and D.~Meloni,
  Phys.\ Lett.\ B {\bf 697}, 477 (2011)
  [arXiv:1010.2329 [hep-ph]].



\bibitem{Hayato:2002sd} 
  Y.~Hayato,
  Nucl.\ Phys.\ Proc.\ Suppl.\  {\bf 112}, 171 (2002).

\bibitem{Gran:2006jn}
  R.~Gran {\it et al.}  [K2K Collaboration],
  Phys.\ Rev.\  D {\bf 74}, 052002 (2006)
  [arXiv:hep-ex/0603034].

\bibitem{:2008eaa}
  A.~Rodriguez {\it et al.}  [K2K Collaboration],
  Phys.\ Rev.\  D {\bf 78}, 032003 (2008)
  [arXiv:0805.0186 [hep-ex]].

\bibitem{AguilarArevalo:2009eb}
  A.~A.~Aguilar-Arevalo {\it et al.}  [MiniBooNE Collaboration],
  Phys.\ Rev.\ Lett.\  {\bf 103}, 081801 (2009)
  [arXiv:0904.3159 [hep-ex]].

\bibitem{AguilarArevalo:2010zc} 
  A.~A.~Aguilar-Arevalo {\it et al.}  [MiniBooNE Collaboration],
  Phys.\ Rev.\ D {\bf 81}, 092005 (2010)
  [arXiv:1002.2680 [hep-ex]].


\bibitem{Kurimoto:2009wq}
  Y.~Kurimoto {\it et al.}  [SciBooNE Collaboration],
  Phys.\ Rev.\  D {\bf 81}, 033004 (2010)
  [arXiv:0910.5768 [hep-ex]].

\bibitem{Nakajima:2010fp}
  Y.~Nakajima {\it et al.}  [SciBooNE Collaboration],
  Phys.\ Rev.\  D {\bf 83}, 012005 (2011)
  [arXiv:1011.2131 [hep-ex]].



\bibitem{Smith:1972xh}
R.~A.~Smith and E.~J.~Moniz,
Nucl.\ Phys.\ B {\bf 43} (1972) 605
[Erratum-ibid.\ B {\bf 101} (1975) 547].


\bibitem{Rein:1980wg}
  D.~Rein and L.~M.~Sehgal,
  Annals Phys.\  {\bf 133}, 79 (1981); 
  Nucl.\ Phys.\  B {\bf 223}, 29 (1983).


\bibitem{Martini:2009uj}
  M.~Martini, M.~Ericson, G.~Chanfray, J.~Marteau,
  Phys.\ Rev.\  {\bf C80}, 065501 (2009)
  [arXiv:0910.2622 [nucl-th]].

\bibitem{Martini:2010ex}
  M.~Martini, M.~Ericson, G.~Chanfray, J.~Marteau,
  Phys.\ Rev.\  {\bf C81}, 045502 (2010)
  [arXiv:1002.4538 [hep-ph]].

\bibitem{AlvarezRuso:2010ia}
  L.~Alvarez-Ruso,
  arXiv:1012.3871 [nucl-th].

\bibitem{Nieves:2011pp} 
  J.~Nieves, I.~Ruiz Simo and M.~J.~Vicente Vacas,
  Phys.\ Rev.\ C {\bf 83}, 045501 (2011)
  [arXiv:1102.2777 [hep-ph]]. 
  J.~Nieves, I.~R.~Simo and M.~J.~V.~Vacas,
  Phys.\ Lett.\ B {\bf 707}, 72 (2012)
  [arXiv:1106.5374 [hep-ph]].

\bibitem{Amaro:2011qb} 
  J.~E.~Amaro, M.~B.~Barbaro, J.~A.~Caballero, T.~W.~Donnelly and J.~M.~Udias,
  Phys.\ Rev.\ D {\bf 84}, 033004 (2011)
  [arXiv:1104.5446 [nucl-th]].


\bibitem{Bodek:2011ps} 
  A.~Bodek and H.~Budd,
  Eur.\ Phys.\ J.\ C {\bf 71}, 1726 (2011)
  [arXiv:1106.0340 [hep-ph]].


\bibitem{Lalakulich:2012ac}
  O.~Lalakulich, K.~Gallmeister and U.~Mosel,
  arXiv:1203.2935 [nucl-th].

\bibitem{Martini:2011ui} 
  M.~Martini,
  arXiv:1110.5895 [hep-ph].



\bibitem{Martini:2011wp} 
  M.~Martini, M.~Ericson and G.~Chanfray,
  Phys.\ Rev.\ C {\bf 84}, 055502 (2011).
  [arXiv:1110.0221 [nucl-th]].


\bibitem{Huber:2004ka}
P.~Huber, M.~Lindner, W.~Winter,
Comput.\ Phys.\ Commun.\  {\bf 167}, 195 (2005).
[hep-ph/0407333];
P.~Huber, J.~Kopp, M.~Lindner, M.~Rolinec, W.~Winter,
Comput.\ Phys.\ Commun.\  {\bf 177}, 432-438 (2007).
[hep-ph/0701187].


\bibitem{Blennow:2009pk} 
M.~Blennow and E.~Fernandez-Martinez,
Comput.\ Phys.\ Commun.\  {\bf 181}, 227 (2010)
[arXiv:0903.3985 [hep-ph]].

\bibitem{Cervera:2000kp}
A.~Cervera, A.~Donini, M.~B.~Gavela, J.~J.~Gomez Cadenas, P.~Hernandez, O.~Mena and S.~Rigolin,
Nucl.\ Phys.\ B {\bf 579} (2000) 17
[Erratum-ibid.\ B {\bf 593} (2001) 731]
[hep-ph/0002108].

\bibitem{Benhar:2009wi} 
  O.~Benhar and D.~Meloni,
  Phys.\ Rev.\ D {\bf 80}, 073003 (2009)
  [arXiv:0903.2329 [hep-ph]].

\bibitem{Leitner:2010kp} 
  T.~Leitner and U.~Mosel,
  Phys.\ Rev.\ C {\bf 81}, 064614 (2010)
  [arXiv:1004.4433 [nucl-th]].

\bibitem{Martini:2012fa} 
  M.~Martini, M.~Ericson and G.~Chanfray,
 Phys.\ Rev.\ D {\bf 85} (2012) 093012
  [arXiv:1202.4745 [hep-ph]].


%
%
%





\bibitem{Donini:2005db}
E.~K.~Akhmedov, R.~Johansson, M.~Lindner, T.~Ohlsson and T.~Schwetz,
JHEP {\bf 0404} (2004) 078
[hep-ph/0402175].

\bibitem{donini2}
A.~Donini, E.~Fernandez-Martinez, D.~Meloni and S.~Rigolin,
Nucl.\ Phys.\ B {\bf 743} (2006) 41
[hep-ph/0512038].


\end{thebibliography}
\end{document}